\documentclass[rmp,twocolumn]{revtex4}

\usepackage{graphicx,amsmath,amssymb,txfonts}

\begin{document}

\title{Role-similarity based functional prediction in networked
  systems:\\ Application to the yeast proteome}

\author{Petter Holme}
\affiliation{Department of Physics, University of Michigan, Ann Arbor,
  MI 48109}
\author{Mikael Huss}
\affiliation{Department of Numerical Analysis and Computer Science,
  Royal Institute of Technology, 100 44 Stockholm, Sweden}

\begin{abstract}
  We propose a general method to predict functions of vertices where:
  1. The wiring of the network is somehow related to the vertex
  functionality. 2. A fraction of the vertices are functionally
  classified. The method is influenced by role-similarity measures of
  social network analysis. The two versions of our prediction scheme
  is tested on model networks were the functions of the vertices are
  designed to match their network surroundings. We also apply these
  methods to the proteome of the yeast \textit{Saccharomyces
    cerevisiae} and find the results compatible with more specialized
  methods.
\end{abstract}

\maketitle

\section{Introduction}

Systems made up of entities that interact pairwise can be modeled as
networks. To comprehend the emergent properties of such systems---the
objective of the study of complex systems and systems biology---one
approach is to investigate the global properties of the corresponding
networks \cite{mejn:rev,ba:rev,harary,wf}. In many cases the
individual entities (or vertices) have distinct functions in the
system. In such cases, provided the wiring of the edges relates to the
function of vertices, one can predict these functions from the
vertices' position in the network. For example, a corporate hierarchy
may be topped by a CEO, followed by a CFO and COO, so a chart of
who reports to whom is enough to identify these positions. Another
problem in this category of much recent interest is to predict protein
functions \cite{hodg:pfp} from the networks of protein interactions
\cite{yook:protein,deng:pfp,hish:pfp,leto:pfp,sama:pfp,vaz:pfp}.
These methods, like other methods based on e.g. protein sequences, 
are important because to confirm a protein function one needs
function-specific and  possibly hard-to-design \textit{in vivo},
genetic or biochemical tests, while interaction and sequence data can
be obtained fairly easily.

In this paper we propose a general method of predicting the functions
of vertices in networked systems where the functions are partly mapped
out. The rationale of our algorithm is to match unknown vertices with
the most similar (judging from the network structure) categorized
vertex and take the functions of the latter vertex as our
forecast. The network similarity concept we ground our method on is
related to the notion of regular equivalence \cite{eve:sim,wf} or role
similarity \cite{regeeco1} of social network theory. Roughly speaking,
two vertices are similar, in this sense, if the network looks alike from
their respective perspectives. We evaluate our method on model
networks where the categories of vertices reflect their placement in
the network. We also apply the method to \textit{S.\ cerevisiae}
protein data obtained from the MIPS data base \cite{pagel:mips} (data
extracted January 23, 2005).

\section{Role similarity and definition of the prediction scheme}

\begin{figure}
  \resizebox*{0.95\linewidth}{!}{\includegraphics{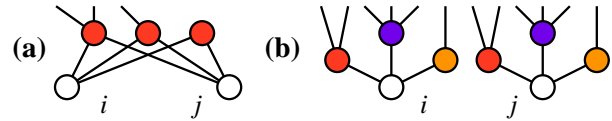}}
  \caption{
    Illustration of structural and regular equivalence. $i$ and $j$
    are structurally equivalent in (a) since they have the same
    neighborhoods, and regularly equivalent in (b) since there is a
    matching of regularly equivalent vertices between the
    neighborhoods. In (b) vertices of the same color are regularly
    equivalent. 
  }
  \label{fig:equ}
\end{figure}

Role similarity refers to rather broad set of concepts and related
measures. Basically, the \textit{role} of a vertex is determined by
the characteristics of the vertices it is connected to
\cite{wf}.\footnote{Note that the nomenclature is somewhat ambiguous. Another
  use of ``role'' is to say that vertices with the similar values of
  vertex-specific structural measures have the same role
  \cite{gui:meta,luss:dolphin}.} Consider
two vertices $i$ and $j$. If their neighborhoods
are similar, we say $i$ and $j$ have high role similarity. The
question how to define the similarity of the neighborhoods $\Gamma_i$
and $\Gamma_j$ leads to two different concepts. One choice matches the
identity of vertices in the neighborhood. This leads to the
\textit{structural equivalence} relation which is true if
$\Gamma_i=\Gamma_j$. Another way to compare neighborhoods is to match
the similarity of vertices in the neighborhood which gives the concept
of \textit{regular equivalence}---if one can pair the vertices of
$\Gamma_i$ with vertices in $\Gamma_j$ such that each pair is
regularly equivalent, then $i$ and $j$ are also regularly
equivalent. Since vertices with the same functions need not, in
general, be close, we will need a similarity score measuring how close
to regular equivalence two vertices are. Following 
Refs.\ \cite{simrank,blondel:sim} we define a similarity score based on
iterating the regular equivalence principle ``two vertices are similar
if they are pointed to, or point to, vertices that themselves
similar.'' In the general case of a directed network with $R$
different types of edges, one implementation of this argument is just
to sum the similarities between vertices of the neighborhoods:
\begin{equation}\label{eq:simdef_i}
  \sigma^\mathrm{I}_{n+1}(i,j) = \sum_{r=1}^R\left[
    \sum_{i'\in\Gamma_{i,r}^{\mathrm{in}}}
    \sum_{j'\in\Gamma_{j,r}^{\mathrm{in}}} \sigma^\mathrm{I}_n (i',j') + 
    \sum_{i'\in\Gamma_{i,r}^{\mathrm{out}}}
    \sum_{j'\in\Gamma_{j,r}^{\mathrm{out}}} \sigma^\mathrm{I}_n
    (i',j')\right],
\end{equation}
where $\sigma^\mathrm{I}_n(i,j)$ is the similarity between $i$ and $j$
after the $n$'th iteration and $\Gamma_{i,r}^{\mathrm{in}}$ is the
in-neighborhood of $i$ with respect to $r$-edges. To avoid
overflow problems we rescale all similarities so that
$\max_{ij}|\sigma^\mathrm{I}_n(i,j)|=S$ after each iteration. We
break the iteration when the sum, before the normalization, has not
changed by more than a $10^{-8}$th of its previous value.

By the Eq.~\ref{eq:simdef_i} definition, high degree vertices will
appear more similar to the average other vertex than low-degree
vertices. To compensate for this effect one may divide by the
appropriate degrees (numbers of neighbors) to obtain:
\begin{widetext}
\begin{equation}\label{eq:simdef_ii}
  \sigma^\mathrm{II}_{n+1}(i,j) = \sum_{r=1}^R\left[
    \frac{1}{k_{i,r}^{\mathrm{in}}\:k_{j,r}^{\mathrm{in}}}
    \sum_{i'\in\Gamma_{i,r}^{\mathrm{in}}}
    \sum_{j'\in\Gamma_{j,r}^{\mathrm{in}}} \sigma^\mathrm{II}_n (i',j') + 
    \frac{1}{k_{i,r}^{\mathrm{out}}\:k_{j,r}^{\mathrm{out}}}
    \sum_{i'\in\Gamma_{i,r}^{\mathrm{out}}}
    \sum_{j'\in\Gamma_{j,r}^{\mathrm{out}}} \sigma^\mathrm{II}_n
    (i',j')\right],
\end{equation}
\end{widetext}
where $k_{i,r}^{\mathrm{in}}$ is the in-degree of $i$ with respect to
$r$-edges. From now on we call $\sigma^\mathrm{I}(i,j)=
\sigma^\mathrm{I}_\infty(i,j)$ of Eq.~\ref{eq:simdef_i} and
$\sigma^\mathrm{II}(i,j)$ of Eq.~\ref{eq:simdef_ii} the I- and
II-similarity between $i$ and $j$ respectively.

As mentioned, we suppose some of the vertices are functionally
categorized. In general we assume one vertex can have many
functions. For pairs of such functionally determined vertices the
above similarities will add no information. Instead we define
a functional similarity
\begin{equation}\label{eq:simdef_f}
  \sigma_f(i,j) = J(F_i,F_j) - \langle J \rangle ,
\end{equation}
for such pairs, where $F_i$ is $i$'s function set (we assume a finite
number of functions) and $J(\:\cdot\:)$ denotes the Jackard index
$J(A,B) = |A\cap B|\:/\:|A\cup B|$ and the average is over all pairs of
categorized vertices. We will later need $\sigma(i,j)=0$ to represent
neutrality which is why we subtract the mean. Whenever a pair of
classified vertices $(i,j)$ appears in the sums of
Eqs.~\ref{eq:simdef_i} or \ref{eq:simdef_ii} we use the
$\sigma_f(i,j)$ value of Eq.~\ref{eq:simdef_f} instead of
$\sigma^\mathrm{I}(i,j)$ or $\sigma^\mathrm{II}(i,j)$. I.e., we assume
the functional classification is more accurate than the
role-similarities and hence do not update the former.

In general we can now define our prediction scheme as follows:
\begin{enumerate}
\item \label{enu:init} For vertex pairs with at least one unclassified
  vertex initialize $\sigma_0(i,j)$ to $0$ if $i\neq j$ and
  to $1 - \langle J \rangle$ otherwise.
\item \label{enu:sim} Calculate the similarity scores for all pairs of
  unique vertices such that at least one is unclassified.
\item \label{enu:choose} For an unclassified vertex $i$, predict the
  function set $F_{\hat{i}}$, where $\hat{i}$ is the classified
  vertex with highest similarity to $i$. If $\hat{i}$ is not unique,
  but a set $\hat{I} = \{\hat{i}_1,\cdots,\hat{i}_m\}$ has the highest
  similarity to $i$, then let the set $G$ of functions present in more
  than half of the set of $j$'s be your guess. If $G$ is empty, let
  $F_j$ for a random $j\in\hat{I}$ be the guess.
\end{enumerate}
The diagonal elements will have maximal functional similarity (which
is why we set them to $1-\langle J \rangle$ in step~\ref{enu:init}),
otherwise we assume neutrality. The backup selection rules in
step~\ref{enu:choose} will typically be needed when unclassified
vertices are structurally equivalent to classified vertices, the use
of the majority rule instead of only a random guess will compensate
for occasional errors in the assignment of functions to classified
proteins. Our parameter $S$ sets the relative importance of the
functional similarities to the subsequent assessments of
$\sigma$. As mentioned above, the functional classification is assumed
to be more accurate than the role-similarities, and it is thus sensible to
choose a $\sigma\in [0,1-\langle J\rangle]$. The appropriate $S$ value
is problem dependent. We will use $S=0{.}8$ which is in this interval
for both our two test cases. To summarize, we have proposed two
versions of our prediction scheme, scheme I and II, corresponding to
I- and II-similarity.

\section{Application to model networks}

To test our prediction algorithm we construct model networks where the
assigned functions of the vertices correspond to their position in the
network. We test the algorithm's size scaling and performance in
sub-ideal conditions by randomly perturbing the network.

\subsection{Definition of the model networks}

\begin{figure*}
  \includegraphics{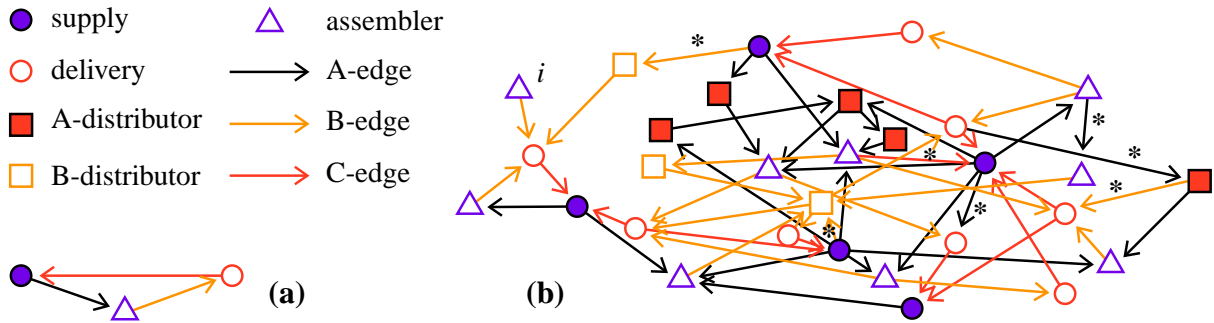}
  \caption{
    Model networks where vertex function and position are related. (a)
    shows the initial network. (b) shows a realization with 30
    vertices and rewiring probability $r=0{.}1$. ``\textbf{*}''
    indicates a rewired edge.
  }
  \label{fig:ill}
\end{figure*}

In defining our model, we will metaphorically use the flow of raw
material, products and information in a manufacturing system. For our
purpose we only need networks where the functions of vertices correspond to
their position in their network surroundings---we will not further
motivate its relevance as a model for manufacturing networks. We
assign five distinct functional classes of the vertices: The
\textit{supply} vertices are the source of the raw material which
flows along \textit{A-edges} to \textit{assembler} vertices. The
assembled products are transported via \textit{B-edges} to
\textit{delivery} vertices that dispatch the products. From the
delivery vertices informational feedback is sent to the supply
vertices  through \textit{C-edges}. Furthermore, the A and B-edges can
fork at \textit{A-} and \textit{B-distributor} vertices.

The precise definition of the model is as follows: Start with the
kernel shown in Fig.~\ref{fig:ill}(a), then grow the network vertex by
vertex. At each iteration, assign, with equal probability, one of the
above functions to the new vertex. Then, depending on the assigned
function, form edges including the new vertex as follows.
\begin{description}
\item[Supply.] Add an A-edge to an assembler or A-distributor, and a
  C-edge from a delivery vertex.
\item[Assembly.] Add an A-edge from an assembler or A-distributor
  vertex, and a B-edge to an assembler or A-distributor.
\item[Delivery.] Add a B-edge from an assembler or B-distributor, and
  a C-edge to a supplier.
\item[A(B)-distribution.] Add an A(B)-edge from an assembler or
  A(B)-distributor vertex, and an A(B)-edge to an assembler or
  A(B)-distributor.
\end{description}
The choice of vertex to attach the new vertex to, given its functional
category, is done with uniform randomness. Note that the number of
edges will on average be twice the number of vertices (two edges are
added per vertex).

From the definition so far, any vertex is identifiable from its
neighborhood---a vertex with incoming C-edges and out-going A-edges is
a supplier, and so on. Real data-sets are seldom perfect---neither in
the wiring of the edges, nor in the functional classification. To test
the prediction scheme under more realistic circumstances we randomize
the network as follows: After generating a network according to the
above scheme, we go through all edges sequentially. With a probability
$r$ detach the from-side of an edge and re-attach it to a randomly
chosen vertex such that no self-edge or multiple edge (of the same
type---A, B or C) is formed. Rewire the to-side likewise with the same
probability. A realization of the algorithm is displayed in
Fig.~\ref{fig:ill}(b). After the rewiring there is not necessarily
enough information to classify a vertex---$i$ in Fig.~\ref{fig:ill}(b)
is an assembler but could just as well have been a B-distributor.

\subsection{Prediction performance}

\begin{figure}
  \resizebox*{\linewidth}{!}{\includegraphics{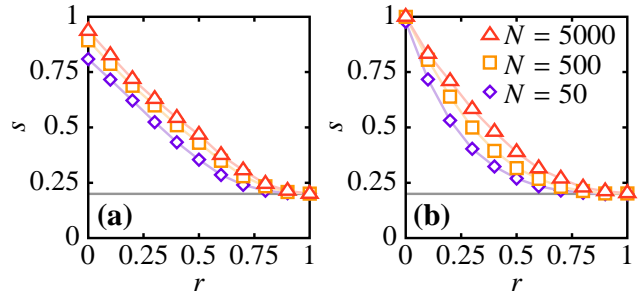}}
  \caption{
    The fraction of correctly predicted functions $s$ for our model
    networks as a function of the rewiring probability $r$.  (a) show
    the results based on I-similarities, (b) is the corresponding plot
    for II-similarities. The points are averaged over $\sim 1000$ runs
    of the network construction and prediction scheme with
    $a=1/50$. Errorbars are smaller than the symbol size. The
    horizontal line marks the limit of random guessing $0{.}2$.
  }
  \label{fig:mod}  
\end{figure}

To test the our prediction scheme we mark a random set of $aN$,
$a\in(0,N)$, vertices unclassified. Then we predict the function of these
vertices and let the average fraction of correctly predicted vertices
$s$ be our performance measure. Fig.~\ref{fig:mod} shows $s$ for
$a=1/50$ and different network sizes, as a function of the the
rewiring probability $r$. In the small-$r$ limit the I-similarity
prediction scheme makes an almost flawless job with $s>99{.}9\%$ for
$N\geqslant 500$. Note, since we have five distinct functions, random
guessing could not do better than $s=1/5$. This value, $s=1/5$, is by
necessity attained in the random limit $r=1$. For small $r$-values the
scheme II performs best, but if $r\lesssim 0{.}2$ scheme I performs
slightly better. The size convergence for scheme I is faster, so in
the large network limit II may outperform I. To understand the
performance of the different schemes we note that scheme I has a
tendency to match an unknown vertex to a known vertex of high
degree. When $r=0$ this effect leads to some mispredictions for scheme
I. But the redundant information about high degree vertices makes the
more robust to minor perturbations, thus the slower decay of the
$s(r)$-curves compared with scheme II.

We observe that the performance increases with the systems size for
both schemes. This is important effect since databases in general grow
in size--our prediction scheme will thus be more accurate with time.
We surmise the explanation lies in, roughly speaking, that the bigger
the network gets, the more likely it is that there is a very good
matching. This is an effect local methods (taking only the surrounding
of a vertex into account) could not utilize. A full explanation of
this effect lies beyond the scope of this paper.

\section{Predicting protein function in yeast}

\begin{figure}
  \resizebox*{0.85\linewidth}{!}{\includegraphics{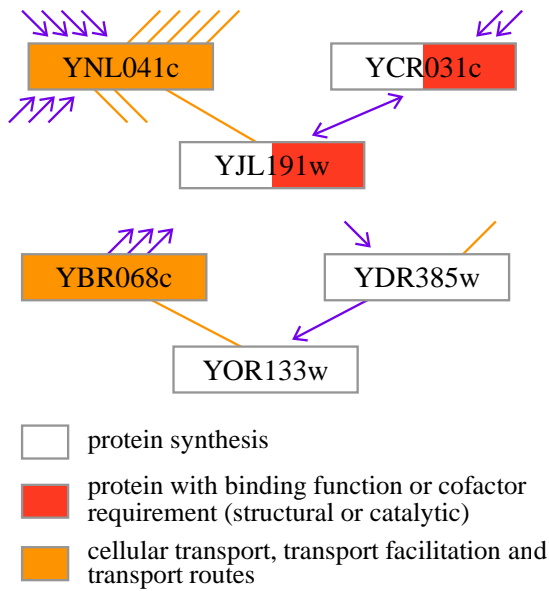}}
  \caption{
    Example from the yeast protein prediction by scheme II on the
    first level functional data. When YJL191w is marked
    unknown it gets matched with YOR133w because their surroundings
    looks similar. The arrowed lines mark genetic regulation edges,
    other lines represent physical interaction.
  }
  \label{fig:pex}
\end{figure}

\subsection{Functional prediction of proteins}

Specifying protein functions experimentally requires demanding and
potentially expensive tests. If one can obtain good guesses of the
functions of an unknown protein, much is gained. During last decade,
there has been a great number of methods suggested for protein
functional prediction, including methods based on  based on sequence
or structure alignments  \cite{paw:seq,irving:struct}, attributes
derived from collections of sequences or
structures \cite{jensen:seq,dobson:struct}, phylogenetic profiles
 \cite{pelle:pfp}, or analysis of protein complexes
\cite{gavin:complexes}.  Much of recent work has concentrated on
functional prediction based on protein-protein interaction data. Many
of these are specialized methods that exploit specific features of
protein-protein interaction data \cite{vaz:pfp,schw:pfp,marc:pfp1,%
marc:pfp2,hodg:pfp,leto:pfp,sama:pfp} (such as that vertices that
interact physically are likely to share some functionality). The more
general approaches \cite{deng:pfp,hish:pfp} are local in the sense
that they are only based on pairwise statistics. For this reason they
may not share the advantageous size scaling properties of our method.

\subsection{Applying the method to protein data}

There are two types of large scale network data available for
\textit{S.\ cerevisiae}: ``physical'' and ``genetic'' protein-protein
interactions. The terms ``physical'' and ``genetic'' refer to the type of
experiment used to deduce the interaction. The genetic experiments
are based on mutation studies, and the evidence from them is of
a more indirect nature. We therefore distinguish
between physical and genetic edges. All edges are undirected. Our data
set, derived from the MIPS data base, has $N=4580$ linked together by
$5129$ genetic regulation edges and $7434$ physical interaction
edges. We removed duplicates, self-edges and interactions where one or
both of the interacting substances were not proteins. The assigned
functions are arranged in a hierarchical fashion, according to the
FunCat categorization scheme \cite{ruepp:funcat} used by the MIPS
database. The first level contains the coarsest description of a
protein's function, such as ``metabolism,'' the second level is more
specified e.g.\ ``amino acid metabolism,'' and so on. We will test our
algorithm of the first and second level of this hierarchy and thus
treat functions that differ in a finer classification as equal. There
are three categories with no substantial functional
information---``ubiquitous expression,'' ``classification not yet
clear-cut'' and ``unclassified proteins.'' We considered vertices with
no other assigned categories than these three uncategorized.

In Fig.~\ref{fig:pex} we show a small example of scheme II in action
on the yeast data. Suppose YJL191w is to be classified (we know it has
the level-1 functions ``protein with binding function \ldots'' and
``protein synthesis''). The classified protein with highest similarity
is YOR133w. This is because YNL041c, which interacts physically with
YJL191w, is functionally identical (at level one of the hierarchy) to
YBR068c that is physically linked to YOR133w. Similarly, YJL191w is
genetically linked with YCR031c, which shares one functional category
with YDR385w, which is genetically linked with YOR133w. These two
features give a high similarity score to the pair YJL191w and YOR133w,
so scheme II guesses that YJL191w has the functional category
``protein synthesis'' but misses the ``protein with binding function
\ldots'' category.

\subsection{Performance of the scheme}

\begin{table}
  \caption{\label{tab:perf} The performance of our methods compared to
    the neighborhood counting method of Ref.\ \cite{schw:pfp}. $s_+$ is
    the average fraction of correct predictions among the predicted
    functions averaged over all the classified proteins. $s_-$ is the
    average fraction of correct predictions among the actual
    functions.}
  \begin{ruledtabular}
    \begin{tabular}{r|cccccc}
      & \multicolumn{3}{c}{level 1} &  \multicolumn{3}{c}{level 2}\\
      & NCM & Scheme I & Scheme II & NCM & Scheme I & Scheme II\\\hline
      $s_+$ & 0{.}269(6) & 0{.}392(6) & 0{.}337(6) &
      0{.}199(5) & 0{.}238(6) & 0{.}220(6) \\
      $s_-$ & 0{.}354(6) & 0{.}291(5) & 0{.}346(7) &
      0{.}252(6) & 0{.}199(5) & 0{.}231(6) \\
    \end{tabular}
  \end{ruledtabular}
\end{table}

For the previously described test networks we know \textit{a priori}
that the number of functions to be predicted is one. The same may be
true for a variety of systems, but not for proteins. With the number
of functions as one variable in the prediction problem we proceed to
replace the success rate $s$ by the two measures \textit{precision}
$s_+$ and \textit{recall} $s_-$ (the names borrowed from corresponding
quantities in the text-mining literature, see e.g.\ Ref.~\cite{rag:tm}
and references therein):
\begin{equation}\label{eq:spm}
  s_+ = \left\langle\frac{n_c}{f_*}\right\rangle \mbox{~and~}
  s_- = \left\langle\frac{n_c}{f}\right\rangle ,
\end{equation}
where $n_c$ is the number of correctly predicted functions, $f$ is the
real number of functions and $f_*$ is the number of predicted
functions. $1-s_+$ is thus the expected fraction of false positive
predictions (and similarly for $s_-$). Both these measures take values
in the interval $[0,1]$ with $0$ meaning that no function is predicted
correctly and $1$ represents perfect prediction. The averages are over
the set of predicted functions in the same kind of leave-one-out
estimates as performed for the test networks.

We follow Refs.\ \cite{vaz:pfp,deng:pfp} and use the neighborhood
counting method (NCM) of Ref.\ \cite{schw:pfp} for reference
values. This method assigns the $f_*$ most frequent functions among
the neighbors of the physical interaction network to the unknown
protein. Considering its simplicity, compared with the more elaborate
procedures listed above, this is a remarkably efficient method. (I.e.,
$f_*$ is a parameter of this model.) In our implementation, if the
$f_*$'th function is not unique we select that randomly. Thus proteins
with no neighbors are assigned $f_*$ functions randomly. Precision and
recall values are displayed in Tab.~\ref{tab:perf}. We use $f_*=2$ for
the NCM which is the closest value to the average number of functions
per protein for both levels one and two in our data set. The values
may look low compared to similar tables in other papers on protein
prediction, but these often do not include low-degree vertices, or use
other performance measures (such as counting the fraction of proteins
with at least one correctly predicted function, and so on). We note
that, like the more disordered test networks, scheme II gives better
performance in general (typically having better recall- but slightly
worse precision-values).

\section{Summary and discussion}

We have proposed methods for predicting the function of vertices in
networked systems where the function of a vertex relates to its
position. The principle behind our scheme is role equivalence as
related to the regular equivalence concept of social network
analysis. I.e., vertices are similar if the network, as seen from the
respective vertices, look similar. We make two extensions to the method
proposed in Refs.\ \cite{simrank,blondel:sim} to networks where some of
the vertices are functionally categorized. The prediction of an
uncategorized protein is then done by copying the functions of the
other vertex with highest role similarity. Our schemes, corresponding
to our two role similarities, are tested on model networks. These are
designed to have a correspondence between the function of the vertex
and their network surrounding. This correspondence can be tuned by a
randomization parameter. We find that the performance of both schemes
increases with the system size (the fraction of unknown vertices and
rewired edges is fixed), which makes the applicability of our methods
increasing with time (as data bases, in general, tend to grow). The
differences between scheme I and II can be described by the fact that,
scheme I gives (compared with scheme II) a higher similarity to
vertex-pairs containing a high-degree vertex. Furthermore, we apply
our method to the \textit{S.\ cerevisiae} proteome. We use the
networks of protein-protein interactions  and obtain results that
compare well with standard methods designed solely with protein
functional prediction in mind. We do not claim that our method
outperform the best specialized protein prediction methods---our aim
is to construct a global method for general functional prediction, and
most protein functional prediction schemes would perform poorly on our
test networks. The ideas of this paper might however contribute to
future, more elaborate, methods for prediction of protein functions.

The basic advantage of our method, as we see it, is that is a very
general method that should apply to functional prediction in many
systems. Moreover, it makes use of global network information,
giving performance that does not decrease as the systems gets
larger. The fact that it is a truly global algorithm---the prediction
of every vertex' functions takes wiring of the whole network into
account---makes it rather slow (compared to e.g.\ specialized protein
functional prediction methods, such as the one proposed in
Ref.\ \cite{schw:pfp}). The execution time scales as $O(M^2)$ (where
$M$ is the total number of edges). But data sets of $10^4$-$10^5$,
which cover e.g.\ the size of proteomes of known organisms, should be
manageable to present day computers. We believe the problem of
functional prediction in different types of networked systems is far
from concluded---both in its full generality and the question how to
utilize the characteristics of more specific systems.

\subsection*{Acknowledgments}

The authors thank Micha Enevoldsen, Elizabeth Leicht and Mark Newman
for comments.

\end{document}